\documentstyle[aps,eqsecnum,preprint,prd,tighten]{revtex}

\def\ov#1{\overline{#1}}

\def\vb#1{\mbox{\boldmath$#1$}}
\def\pd#1#2{\frac{\partial #1}{\partial #2}}
\def\wh#1{\widehat{#1}}
\def\bdot{\,\vb{\cdot}\,}
\def\btimes{\,\vb{\times}\,}

\begin{document}
\preprint{LBNL-45042 \hfill UCB-PTH-99/39}

\title{Magnetic Field Generation from Self-Consistent Collective 
Neutrino-Plasma Interactions% 
\thanks{This work was performed under Department of Energy Contracts 
No.~PDDEFG-03-95ER-40936 and DE-AC03-76SF00098, in part by the 
National Science Foundation under grant PHY-95-14797, and in part also 
by Alfred P. Sloan Foundation.}
}

\author{A.J.~Brizard, H.~Murayama and J.S.~Wurtele}
\address{Department of Physics, University of California, Berkeley,
California 94720\\ {\rm and}\\ 
Lawrence Berkeley National Laboratory, Berkeley, California 94720}

\date{\today}
\maketitle
\vfill
 
\begin{abstract}
  A new Lagrangian formalism for self-consistent collective neutrino-plasma
  interactions is presented in which each neutrino species is
  described as a classical ideal fluid. The neutrino-plasma fluid equations
  are derived from a covariant relativistic variational principle in
  which finite-temperature effects are retained. This new formalism is
  then used to investigate the generation of magnetic fields and the
  production of magnetic helicity as a result of collective neutrino-plasma 
  interactions.
\end{abstract}


\section{INTRODUCTION}
\label{sec:intro}

Photons, neutrinos and plasmas are ubiquitous in the universe 
\cite{KT,Peebles}. During the early universe, it is expected that photons and 
neutrinos interacted quite strongly with hot primordial plasmas 
\cite{primordial}. Although photons and neutrinos decoupled from plasmas 
relatively early after the big bang \cite{KT,Peebles}, there are still 
conditions today where neutrino-plasma interactions might be important. 
For example, during a supernova explosion \cite{ST,CHB,Cooperstein}, intense 
neutrino fluxes are generated as result of the gravitational collapse of the 
stellar core. It is generally believed that the outgoing neutrino flux needs to 
transfer energy and momentum to the surrounding plasma in order to produce the 
observed explosion. 

The self-consistent collective interaction between photons and plasmas is
traditionally treated classically (i.e., without quantum-mechanical 
effects), where plasma particles are either treated within a fluid or a kinetic 
picture, while photons are described in terms of an electromagnetic 
field. For a self-consistent treatment of collective electromagnetic-plasma 
interactions (see Ref.~\cite{Brizard}, for example), one considers both the 
influence of electromagnetic fields on plasma dynamics and the generation of 
electromagnetic fields by plasma currents. The interaction between photons and
neutrinos, on the other hand, requires a full quantum-mechanical treatment and 
has been the subject of recent interest \cite{photon_neutrino}.

Neutrino-plasma interactions involve charged and neutral currents associated
with the weak force \cite{Taylor,electroweak} (through the exchange of
$W^{\pm}$ and $Z^{0}$ bosons, respectively). The collective interactions
studied here apply to the intense neutrino fluxes.  Discrete (i.e.,
collisional) neutrino-plasma interactions, on the other hand, involve
scattering of individual particles; such discrete neutrino-plasma particle 
effects will be omitted in the present work.

The purpose of the present work is to investigate the self-consistent 
collective interaction between neutrinos and plasmas in the presence of 
electromagnetic fields. The inclusion of electromagnetic effects is a departure 
from conventional hydrodynamic models used in investigating neutrino 
interactions with astrophysical plasmas \cite{CHB}. Here, we investigate the 
collective processes 
\begin{equation}
EM \;\rightarrow\; \sigma \;\rightarrow\; \nu 
\label{eq:forward}
\end{equation}
and 
\begin{equation}
\nu \;\rightarrow\; \sigma \;\rightarrow\; EM.
\label{eq:reverse}
\end{equation}
In the first process, the neutrino ($\nu$) dynamics is influenced by an 
electromagnetic field ($EM$) with a plasma ($\sigma$) background acting as an 
intermediary, even though neutrinos are chargeless particles. In the second 
process, electromagnetic fields are generated as a result of plasma currents 
produced by neutrino ponderomotive effects. The problem of magnetic-field 
generation associated with self-consistent collective neutrino-plasma 
interactions is thus investigated here within the context of the process 
(\ref{eq:reverse}).

\subsection{Notation}

In the present paper, the Latin subscript $s$ refers to different components of
the neutrino-plasma fluid: the subscript $s=\nu$ refers to neutrinos while the
subscript $s = \sigma$ refers to components of the plasma other than photons
and neutrinos. To avoid confusion, we use the Greek letters $\alpha, \beta,
\cdots$ for Lorentz indices rather than traditional $\mu, \nu, \cdots$; for
example, the flux four-vector is $J^{\alpha} = N u^{\alpha}$, with proper 
density $N$ (Lorentz scalar) and normalized four-velocity $u^{\alpha} = (u^{0},
{\bf u})$. In certain cases, objects with Lorentz indices may not be covariant; 
for instance, the fluid velocity $v^{\alpha} = u^{\alpha}/u^{0}$ is not 
covariant and the number density in a given frame $n = Nu^{0}$ is not a
Lorentz scalar.   The symbols in bold face are three-vectors while those in 
Sans Serif are four-dimensional tensors (such as ${\sf F}$ for the 
electromagnetic field strength $F_{\alpha\beta}$). The dot $\cdot$ describes 
the contraction of a Lorentz index or an inner product of two three-vectors if 
in bold face. Here, we employ the metric $g_{\alpha\beta} = {\rm diag}(1,-1,-1,
-1)$ and, hence, $a\cdot b \equiv a^{0}b^{0} - {\bf a}\bdot{\bf b}$.

\subsection{Neutrino Descriptions for Collective Neutrino-Plasma 
Interactions}  

To study collective neutrino-plasma interactions, neutrinos can either be 
described in terms of Dirac spinor fields \cite{Taylor,electroweak,PP,NR,EC}, 
Klein-Gordon scalar fields \cite{Bingham,Brizard_Wurtele}, classical
non-relavistic fluids \cite{Tajima}, or relativistic quasi-particles 
\cite{Silva,Silva_prl}. In all these descriptions, the interaction between
neutrinos (of type $\nu$) and plasma particles (of species $\sigma$) is 
described in terms of an effective weak-interaction charge $G_{\sigma\nu}$.  In 
general, $G_{\sigma\nu}$ has the following property \cite{PP}:
\begin{equation}
G_{\sigma\nu} \;=\; -\,G_{\ov{\sigma}\nu} \;=\; -\,G_{\sigma\ov{\nu}} \;=\;
G_{\ov{\sigma}\ov{\nu}},
\label{eq:identity_G}
\end{equation}
where $\sigma$ ($\ov{\sigma}$) denotes a matter (anti-matter) species and $\nu$
($\ov{\nu}$) denotes a neutrino (anti-neutrino) species. The effective charge
$G_{\sigma\nu}$ depends on the Fermi weak-interaction constant $G_{F} \; 
(\approx 9 \times 10^{-38}$ eV cm$^{3}$), the Weinberg angle $\theta_{W}$ 
($\sin^{2}\theta_{W} \approx 0.23$ \cite{electroweak}), and the species 
$\sigma$ and $\nu$. For example, for neutrinos interacting with unpolarized 
electrons ($e$), protons ($p$) and neutrons ($n$), one finds \cite{PP}  
\begin{equation}
G_{\sigma\nu} \;=\; \sqrt{2}G_{F}\; 
\left[\; \delta_{\sigma e} \delta_{\nu \nu_{e}}\;+\; \left( 
I_{\sigma} \;-\; 2Q_{\sigma}\,\sin^{2}\theta_{W} \right) \;\right],
\label{eq:def_G}
\end{equation}
where $I_{\sigma}$ is the weak isotopic spin for particle species $\sigma$ 
($I_{e} = I_{n} = -1/2$ and $I_{p} = 1/2$) and $Q_{\sigma} 
\equiv q_{\sigma}/e$ is the normalized electric charge. Here, the first term 
in (\ref{eq:def_G}) is due to charged weak currents (and thus applies only to 
electrons and electron-neutrinos), while the remaining terms are due to neutral 
weak currents (and thus apply to all species). 

To assist us in investigating self-consistent collective neutrino-plasma
interactions in the present work, all neutrino and particle species are treated 
as ideal classical fluids. For this purpose, we proceed with the classical 
fluid limit for plasma-particles in the Dirac description expressed in terms 
of the correspondence 
\begin{equation}
\ov{\psi}_{\sigma}\left( \wh{\gamma}^{\alpha}/c \right)\psi_{\sigma} 
\;\rightarrow\; J_{\sigma}^{\alpha} \;\equiv\; (n_{\sigma},\vb{J}_{\sigma}), 
\label{eq:corr}
\end{equation}
where $\psi_{\sigma}$ is the Dirac spinor field for particle species $\sigma$ 
(with $\wh{\gamma}^{\alpha}$ denoting Dirac matrices) while $n_{\sigma}$ and 
$\vb{J}_{\sigma} \equiv n_{\sigma} {\bf v}_{\sigma}/c$ are the particle density 
and (normalized) particle flux for each plasma-fluid species $\sigma$ in the 
lab reference frame, respectively. In this limit, the propagation of a 
neutrino test-particle of type $\nu$ in a background plasma is determined by 
the effective potential \cite{potential}
\begin{equation}
V_{\nu}({\bf x},{\bf v},t) \;\equiv\; \sum_{\sigma} G_{\sigma\nu}\,
\left[\; n_{\sigma}({\bf x},t) \;-\; \vb{J}_{\sigma}({\bf x},t)\bdot
\frac{{\bf v}}{c} \;\right],
\label{eq:pot_nu}
\end{equation}
where $({\bf x},{\bf v})$ denote the neutrino's position and velocity. We note 
that neutrino propagation in matter is a topic at the heart of the problem of 
neutrino oscillations in matter \cite{MSW,Bethe,mode_conv} and the solar 
neutrino problem \cite{Balantekin}. Although the term $\vb{J}_{\sigma}\bdot
{\bf v}/c$ is a relativistic correction to $n_{\sigma}$ in (\ref{eq:pot_nu}),
we keep it for the following reason. For a primordial plasma with a single 
family of particles ($s = \sigma$) and anti-particles ($s = \ov{\sigma}$), we 
find from (\ref{eq:identity_G}) 
\begin{equation}
\left. \begin{array}{rcl}
\sum_{s = \sigma,\ov{\sigma}}\; G_{s\nu}\,n_{s} & = & 0 \\
 &   & \\
\sum_{s = \sigma,\ov{\sigma}}\; G_{s\nu}\,\vb{J}_{s} & = & G_{\sigma\nu}\;
\left( \vb{J}_{\sigma} - \vb{J}_{\ov{\sigma}} \right)
\end{array} \right\},
\label{eq:identity_PG}
\end{equation}
and thus the effective neutrino potential (\ref{eq:pot_nu}) becomes $V_{\nu} = 
G_{\sigma\nu}\,(\vb{J}_{\ov{\sigma}} - \vb{J}_{\sigma})\bdot{\bf v}/c$, for 
each $(\sigma,\ov{\sigma})$-family. Hence, keeping this relativistic correction 
is necessary for the description of collective neutrino interactions with a 
primordial plasma \cite{hot}. The model presented here therefore retains all
relativistic effects associated with the neutrino and plasma fluids. 

For a self-consistent description of collective neutrino-plasma
interactions in which neutrino ponderomotive effects on the background medium
are included, we now use a similar classical-fluid correspondence for 
the neutrinos. The propagation of a plasma test-particle of species $\sigma$ 
(with electric charge $q_{\sigma}$) in a background medium composed of a 
neutrino fluid of type $\nu$ and an electromagnetic field is determined by the 
potential 
\begin{equation}
V_{\sigma}({\bf x},{\bf v},t) \;\equiv\; \left[ q_{\sigma}\,\phi({\bf x},t) + 
\sum_{\nu}\, G_{\sigma\nu}\,n_{\nu}({\bf x},t) \right] \;-\; \left[ q_{\sigma}\,
{\bf A}({\bf x},t) + \sum_{\nu}\,G_{\sigma\nu}\,\vb{J}_{\nu}({\bf x},t) 
\right]\bdot \frac{{\bf v}}{c}, 
\label{eq:pot_sigma}
\end{equation}
where $n_{\nu}$ and $\vb{J}_{\nu} \equiv n_{\nu}{\bf v}_{\nu}/c$ are the 
neutrino density and (normalized) neutrino flux in the lab reference frame, 
respectively, $\phi$ and ${\bf A}$ are the electromagnetic potentials, and
$({\bf x},{\bf v})$ denote the plasma-particle's position and velocity. It is
interesting to note how the right side of (\ref{eq:pot_sigma}) links the 
electrostatic scalar potential $\phi$ and the neutrino density $n_{\nu}$, on 
the one hand, and the magnetic vector potential ${\bf A}$ and the neutrino flux 
vector $\vb{J}_{\nu}$, on the other hand. We will henceforth refer to the
approximation whereby $\vb{J}_{\sigma}$ and $\vb{J}_{\nu}$ are omitted in
(\ref{eq:pot_nu}) and (\ref{eq:pot_sigma}) as the {\it weak}-electrostatic
(or non-relativistic) approximation.
   
Although we assume that each neutrino flavor has a finite mass, this assumption 
is not crucial to the development of our model; see Section \ref{sec:lagrangian}
for a discussion of neutrino-fluid dynamics for arbitrary neutrino masses. 
Furthermore we shall ignore all quantum mechanical effects, including effects 
due to strong magnetic fields \cite{magnetic} (i.e., we assume $B/B_{{\rm QM}} 
\equiv \hbar\Omega_{e}/m_{e} c^{2} \ll 1$, where $\Omega_{e} \equiv eB/m_{e}c$ 
is the electron gyrofrequency and $B_{{\rm QM}} \sim 4\times 10^{13}$ G). 
Hence, although magnetic fields appear explicitly in our model, they are not 
considered strong enough to modify the form of the interaction potentials 
(\ref{eq:pot_nu}) and (\ref{eq:pot_sigma}).

\subsection{Magnetic-Field Generation due to Neutrino-Plasma Interactions}

An important application of the process (\ref{eq:reverse}) involves the 
prospect of generating magnetic fields in an 
unmagnetized plasma as a result of collective neutrino-plasma interactions. 
This application may be of importance in investigating magnetogenesis in the 
early universe (e.g., see Ref.~\cite{Peebles}). A similar process of 
magnetic-field generation occurs in laser-plasma interactions whereby an 
intense laser pulse propagating in a nonuniform plasma generates a quasi-static 
magnetic field. This process was first studied theoretically 
\cite{GMA,Haines,MT} and was recently confirmed experimentally
\cite{B_generation}.

The generation of magnetic fields by collective neutrino-plasma interactions 
was first contemplated in the non-relativistic (weak-electrostatic) limit by 
Shukla {\it et al.} \cite{Shukla_97,Shukla_98}. The covariant (relativistic) 
Lagrangian approach introduced by Brizard and Wurtele \cite{Brizard_Wurtele}, 
however, revealed the presence of additional ponderomotive terms missing from 
previous analysis \cite{Bingham,Shukla_97,Shukla_98}. These additional 
ponderomotive terms involve the time derivative of the neutrino flux 
$\partial_{t}\vb{J}_{\nu}$ and the curl of the neutrino flux $\nabla\btimes
\vb{J}_{\nu}$ (henceforth referred to as the neutrino-flux vorticity), which 
are shown here to lead to significantly different predictions 
regarding neutrino-induced magnetic-field generation. In fact, we show that 
magnetic-field generation due to neutrino-plasma interactions is not possible 
without these new terms.

\subsection{Organization}

The remainder of this paper is organized as follows. In Section 
\ref{sec:lagrangian}, the Lagrangian formalism for ideal fluids is introduced.
In Section \ref{sec:constrained}, a variational principle for collective 
neutrino-plasma interactions in the presence of an electromagnetic field is 
presented. This Lagrangian formalism is fully relativistic and covariant and 
can thus be generalized to include general relativistic effects (e.g., see 
Refs.~\cite{Brown,Achterberg}).  In Section \ref{sec:self}, the nonlinear 
neutrino-plasma fluid equations and the Maxwell equations for the 
electromagnetic field are derived. Through the Noether method 
\cite{Whitham,Similon,Mills}, an exact energy-momentum conservation law is also 
derived and the process of energy-momentum transfer from the neutrinos to the 
electromagnetic field and the plasma is discussed. In Section 
\ref{sec:magnetic}, magnetic-field generation, magnetic-helicity production
and magnetic equilibrium involving neutrino-plasma interactions are 
investigated. Here, we find that neutrino-flux vorticity ($\nabla \btimes 
\vb{J}_{\nu}$) plays a fundamental role in all three processes.  We summarize 
our work in Section \ref{sec:summary} and discuss future work.

\section{LAGRANGIAN DENSITY FOR A FREE IDEAL FLUID}
\label{sec:lagrangian}

The present Section is dedicated to the derivation of a suitable Lagrangian
density for a free ideal fluid from an existing single-particle Lagrangian for 
a free particle of arbitrary mass (including zero). The difficulty with dealing 
with the case of free neutrinos as particles is that their mass may be zero. 
Since the relativistic Lagrangian $L$ for a free single particle of mass $m$ is 
\cite{Goldstein}
\begin{equation}
L \;=\; -\,mc^{2}\,\gamma^{-1} \;\equiv\; -\, mc \left( \frac{dx^{\alpha}}{dt}
\frac{dx_{\alpha}}{dt} \right)^{1/2},
\label{eq:class_lag}
\end{equation}
it is not obvious how to handle the limiting case of zero mass. This difficulty 
is resolved in \cite{string_theory} as follows (see also Ref.~\cite{Peebles}).

\subsection{Single-Particle Lagrangian}

Consider the primitive Lagrangian
\begin{equation}
L_{p} \;=\; {\bf p}\bdot\dot{{\bf x}} - E\,\dot{t} \;\equiv\; -\, p_{\alpha}c
\,v^{\alpha}
\label{eq:prim_lag}
\end{equation}
for a particle of arbitrary rest-mass $m$ (including zero), where $(x,p)$ are
coordinates in the eight-dimensional phase space in which the particle moves and
$\dot{x}^{\alpha} = (c,{\bf v}) \equiv cv^{\alpha}$. Although the particle's 
space-time location $x^{\alpha} = (ct,{\bf x})$ is arbitrary, its four-momentum 
$p^{\alpha} = (E/c,{\bf p})$ is not since the particle's physical motion is 
constrained to occur on the mass shell 
\begin{equation}
p_{\alpha}p^{\alpha} \;=\; m^{2}c^{2}. 
\label{eq:mass_shell}
\end{equation}
Here, $u^{\alpha} \equiv \gamma v^{\alpha}$ is the normalized four-velocity and 
$\gamma = (1 + |{\bf u}|^{2})^{1/2}$ is the relativistic factor.

Since the mass constraint (\ref{eq:mass_shell}) cannot be derived from the 
primitive Lagrangian (\ref{eq:prim_lag}), we explicitly introduce it by means 
of a Lagrange multiplier:
\begin{equation}
L_{p} \;\equiv\; -\;p_{\alpha}c\,v^{\alpha} \;-\; \frac{1}{2\lambda}\left(\,
m^{2}c^{4} - p_{\alpha}p^{\alpha}\,c^{2} \,\right),
\label{eq:lag_mult}
\end{equation}
where $\lambda^{-1}$ is the Lagrangian multiplier  and the factor $1/2$ is 
added for convenience. Since the Lagrangian (\ref{eq:lag_mult}) is independent 
of $\dot{p}_{\alpha}$, the Euler-Lagrange equation for $p_{\alpha}$ yields
\begin{equation}
\pd{L_{p}}{p_{\alpha}} \;=\; -\;cv^{\alpha} \;+\;
\frac{p^{\alpha}c^{2}}{\lambda} \;\equiv\; 0,
\label{eq:EL_p}
\end{equation}
from which we obtain 
\begin{equation}
p^{\alpha} = \lambda\,v^{\alpha}/c.
\label{eq:p_eq}
\end{equation}
Using the mass constraint (\ref{eq:mass_shell}) and the identity $v\cdot v
\equiv \gamma^{-2}$, the relation (\ref{eq:p_eq}) yields
\begin{equation}
\lambda \;=\; \gamma mc^{2},
\label{eq:lambda_eq}
\end{equation}
i.e., $\lambda$ is the energy of a single particle of mass $m$.

If we now substitute (\ref{eq:p_eq}) into the primitive Lagrangian 
(\ref{eq:lag_mult}) (i.e., by constraining the physical motion to take place on 
the mass shell), we find the physical Lagrangian
\begin{equation}
L(v;\lambda) \;\equiv\; L_{p}(x;p = \lambda v/c;\lambda) \;=\; -\; 
\frac{m^{2}c^{4}}{2\lambda} \;-\; \frac{\lambda}{2\gamma^{2}}.
\label{eq:phys_lag}
\end{equation}
This Lagrangian now depends only on $v^{\alpha}$ and $\lambda$ (for a free
particle, there is no space-time dependence in the Lagrangian). The 
Euler-Lagrange equation for $\lambda$ now yields
\begin{equation}
\pd{L}{\lambda} \;=\; \frac{1}{2} \left( \frac{m^{2}c^{4}}{\lambda^{2}} \;-\;
\frac{1}{\gamma^{2}} \right) \;\equiv\; 0,
\label{eq:EL_lambda}
\end{equation}
which gives (\ref{eq:lambda_eq}). Substituting of (\ref{eq:lambda_eq}) into 
(\ref{eq:phys_lag}) yields the standard Lagrangian (\ref{eq:class_lag}).

For a massless particle, on the other hand, the condition (\ref{eq:EL_lambda}) 
yields 
\begin{equation}
\gamma^{-2} \;=\; v_{\alpha}\,v^{\alpha} \;\equiv\; 0, 
\label{eq:EL_massless}
\end{equation}
which states that massless particles travel at the speed of light. Here, 
$\lambda$ is still the massless particle's energy since (\ref{eq:p_eq}) gives
$p^{0} \equiv \lambda/c$.  For a massless particle, the single-particle 
Lagrangian is therefore simply given by the last term in (\ref{eq:phys_lag}), 
i.e.,
\begin{equation}
L(v;\lambda) \;\equiv\; -\,\frac{\lambda}{2}\,v_{\alpha}\,v^{\alpha}.
\label{eq:massless_lag}
\end{equation}
This Lagrangian appears in the bosonic part of the 
Lagrangian for a spinning particle \cite{string_theory}.  The Lagrange 
multiplier $\lambda^{-1}$ corresponds to the ``einbein'' which 
describes the square-root metric $e=\sqrt{g}$ along the world line in 
a particular gauge where the world line is parameterized by time.

\subsection{Lagrangian density for a Free Ideal Fluid}

We now discuss the passage from the finite-dimensional single-particle
Lagrangian formalism based on (\ref{eq:class_lag}) to an infinite-dimensional 
fluid Lagrangian formalism. To obtain a Lagrangian density for a fluid composed 
of such particles, we multiply (\ref{eq:class_lag}) by the reference-frame 
density $n$, noting that the proper density is $N \equiv n \gamma^{-1}$.
The Lagrangian for a cold ideal fluid is therefore
\begin{equation}
        {\cal L}_{0} = - mc^{2} N
                = -mc^{2} n \sqrt{v^{\alpha} v_{\alpha}}
                = -mc^{2} \sqrt{J^{\alpha} J_{\alpha}},
\end{equation}
where $J^{\alpha} = n v^{\alpha} = \langle \bar{\psi} \wh{\gamma}^{\alpha} 
\psi/c \rangle$ is the flux four-vector with a suitable ensemble average 
$\langle \cdots \rangle$.  The Lorentz invariance is manifest in the last 
expression.  

Another contribution to the Lagrangian density of an ideal fluid
is the term $-\,N\epsilon(N,S)$ associated with the internal energy density of 
the fluid in its rest frame, where the internal energy $\epsilon(N, S)$ is a 
function of the proper fluid density $N$ and its entropy $S$ (a 
Lorentz scalar).  By combining these two terms, the Lagrangian density 
for a free relativistic fluid is therefore written as
\begin{equation}
{\cal L}_{0} \;=\; -\,N \left[\; mc^{2} \;+\; \epsilon(N,S) \;\right]
\;\equiv\; -\,N\,\varepsilon(N,S),
\label{eq:free_lag_d}
\end{equation}
where the total internal energy
\begin{equation}
\varepsilon(N,S) \;\equiv\; mc^{2} \;+\; \epsilon(N,S)
\label{eq:varepsilon}
\end{equation}
includes the particle's rest energy.  

As discussed above, the single-particle Lagrangian for a free massless
particle is given as (\ref{eq:massless_lag}). The Lagrangian density for a
cold ideal fluid composed of massless neutrinos is therefore given as
\begin{equation}
{\cal L}_{0} \;\equiv\; -\,\frac{\lambda'_{\nu}}{2}\;J_{\nu}\cdot J_{\nu} 
= -\,\frac{n_{\nu}\lambda_{\nu}}{2}\;v_{\nu}\cdot v_{\nu}, 
\label{eq:massless_lag_d}
\end{equation}
where $\lambda'_{\nu}$ is a Lorentz-scalar Lagrange multiplier field.  The last 
expression is equivalent upon changing the variable $\lambda_{\nu} = n_{\nu} 
\lambda'_{\nu}$.  

\section{CONSTRAINED VARIATIONAL PRINCIPLE}
\label{sec:constrained}

The self-consistent nonlinear neutrino-plasma fluid equations presented in this
paper are derived from the variational principle:
\begin{equation}
\delta \,\int d^{4}x\;{\cal L}\left(A^{\alpha},F^{\alpha\beta};\,N_{s},
u_{s}^{\alpha}, S_{s}\right) \;=\; 0,
\label{eq:vp}
\end{equation}
where in addition to its dependence on the electromagnetic four-potential 
$A^{\alpha}$ and the Faraday tensor $F^{\alpha\beta}$, the Lagrangian density
${\cal L}$ depends on the proper density $N_{s} \equiv n_{s}\gamma_{s}^{-1}$, 
the normalized fluid four-velocity $u_{s}^{\alpha} \equiv (\gamma_{s},
{\bf u}_{s})$, and the proper internal energy (per particle) $\varepsilon_{s}$ 
for each fluid species $s$ (here, $s = \sigma$ denotes a plasma-fluid species 
and $s = \nu$ denotes a neutrino-fluid species). 

The proper internal energy $\varepsilon_{s}(N_{s}, S_{s})$ includes 
the particle's rest energy [see Eq.~(\ref{eq:free_lag_d})] and depends 
on the proper density $N_{s}$ and the entropy $S_{s}$ (a Lorentz 
scalar).  The first law of thermodynamics \cite{Tolman,Weinberg,MTW} 
is written as
\begin{equation}
d\varepsilon_{s} \;=\; T_{s}\, dS_{s} \;-\; p_{s}\, dN_{s}^{-1}, 
\label{eq:first_law}
\end{equation}
where $T_{s}$ is the proper temperature and $p_{s}$ is the scalar pressure for 
fluid species $s$. In what follows we use the chemical potential for each fluid 
species $s$: 
\begin{equation}
\mu_{s} \;\equiv\; \partial(\varepsilon_{s} N_{s})/\partial N_{s} \;=\; 
\varepsilon_{s} \;+\; p_{s}/N_{s},
\label{eq:chem_pot}
\end{equation} 
which represents the total energy required to create a particle of species $s$ 
and inject it in a fluid sample composed of particles of the same species.
Associated with the definition for the chemical potential (\ref{eq:chem_pot}),
we also use the identity
\begin{equation}
\partial^{\alpha}\mu_{s} \;=\; T_{s}\,\partial^{\alpha}S_{s} \;+\; N_{s}^{-1}\,
\partial^{\alpha}p_{s}.
\label{eq:grad_mu}
\end{equation}
Note that the independent fluid variables for each fluid species are $N_{s}$,
$u_{s}^{\alpha}$ and $S_{s}$ although other combinations are possible 
\cite{Brown}.

The Lagrangian formulation for the nonlinear interaction between neutrino and 
plasma fluids in the presence of an electromagnetic field is expressed in terms 
of the Lagrangian density
\begin{equation}
{\cal L} \;=\; -\,\sum_{s=\sigma,\nu}\,N_{s}\,\varepsilon_{s} \;-\; 
\sum_{\sigma}\,J_{\sigma}\;\cdot\; \left( q_{\sigma}\,A + \sum_{\nu}\, 
G_{\sigma\nu}\, J_{\nu} \right) \;+\; \frac{1}{16\pi}\, {\sf F}:{\sf F},
\label{eq:lag_d}
\end{equation}
where ${\sf F}: {\sf F} \equiv F^{\alpha\beta}\, F_{\beta\alpha}$.
The first term in (\ref{eq:lag_d}) denotes the total internal energy density of fluid $s$. The second term denotes the standard coupling between a charged 
(plasma) fluid and an electromagnetic field. The third term denotes the 
coupling between the neutrino-fluid species $\nu$ and the plasma-fluid species 
$\sigma$. Note that the second and third terms can be written as 
$\sum_{\sigma}\,n_{\sigma} V_{\sigma}$, where the single-particle velocity 
${\bf v}$ in (\ref{eq:pot_sigma}) is replaced with the fluid velocity 
${\bf v}_{\sigma}$.   The fourth term is the familiar electromagnetic field 
Lagrangian.

In the variational principle (\ref{eq:vp}), the variation $\delta{\cal L}$ is 
explicitly written as
\begin{equation}
\delta{\cal L} \;\equiv\; \delta A\cdot \pd{{\cal L}}{A} \;-\; \delta {\sf F} :
\pd{{\cal L}}{{\sf F}} \;+\; \sum_{s} \left( \delta N_{s}\;\pd{{\cal L}}{N_{s}}
\;+\; \delta u_{s}\cdot \pd{{\cal L}}{u_{s}} \;+\; \delta S_{s}\;
\pd{{\cal L}}{S_{s}} \right),
\label{eq:delta_l}
\end{equation}
where $\delta F_{\alpha\beta} = \partial_{\alpha}\delta A_{\beta} - 
\partial_{\beta} \delta A_{\alpha}$ so that the second term in 
(\ref{eq:delta_l}) can also be written as $+2\,\partial\delta A:\partial
{\cal L}/\partial{\sf F}$.  In constrast to other variational principles 
\cite{Brown,SW}, the Eulerian variations $\delta N_{s}$, $\delta u_{s}$ and 
$\delta S_{s}$ in (\ref{eq:delta_l}) are not arbitrary but are instead 
{\it constrained}\/.

To obtain the correct Eulerian variation, recall that the variation of 
the fluid motion is an infinitesimal displacement of the fluid 
elements.  With a fluid element $s$ described by the four-coordinate 
$x_{s}^{\alpha}$, the normalized velocity four-vector is given by
\begin{equation}
        u_{s}^{\alpha} (x) = \frac{d x_{s}^{\alpha}}{d\tau} 
        \left( 
        \frac{dx_{s}}{d\tau}\cdot\frac{dx_{s}}{d\tau}\right)^{-1/2} \;\equiv\;
        \left| \frac{dx_{s}}{d\tau}\right|^{-1}\,\frac{dx_{s}^{\alpha}}{d\tau},
\end{equation}
where $\tau$ parametrizes the world line of the fluid element.  
Under the infinitesimal displacement $x_{s}^{\alpha} \rightarrow 
x_{s}^{\alpha} + \delta\xi_{s}^{\alpha}$ [with $\delta\xi_{s}^{\alpha} \equiv 
(\delta\xi_{s}^{0},\delta\vb{\xi}_{s})$], the apparent variation at a 
position following a fluid element along its worldline is
\begin{eqnarray}
        \lefteqn{
        \frac{d \delta \xi_{s}^{\alpha}}{d\tau}\;
        \left| \frac{dx_{s}}{d\tau}\right|^{-1} 
        - \frac{d x_{s}^{\alpha}}{d\tau} 
        \left| \frac{dx_{s}}{d\tau}\right|^{-3}
        \frac{d \delta\xi_{s}}{d\tau}\cdot\frac{dx_{s}}{d\tau}
        } \nonumber \\
        & & \qquad  = (u_{s} \cdot \partial) \delta \xi_{s}^{\alpha}
        - u_{s}^{\alpha} \;[u_{s\beta} 
                (u_{s}\cdot \partial) \delta \xi_{s}^{\beta}]
        \equiv h_{s}^{\alpha\beta} (u_{s} \cdot \partial) \delta \xi_{s\beta},
\end{eqnarray}
with $u_{s}\cdot\partial \equiv |dx_{s}/d\tau|^{-1}\;d/d\tau$ and
\begin{equation}
h_{s}^{\alpha\beta} \;\equiv\; g^{\alpha\beta} \;-\; u_{s}^{\alpha}u_{s}^{\beta}
\label{eq:h}
\end{equation}
is a symmetric projection tensor \cite{Weinberg} (i.e., ${\sf h}_{s}\cdot u_{s} 
\equiv 0$).  The Eulerian variation at a fixed space-time location is 
therefore given by \cite{Achterberg}
\begin{equation}
        \delta u_{s}^{\alpha} (x) = h_{s}^{\alpha\beta} (u_{s} \cdot
        \partial) \delta \xi_{s\beta}
        - (\delta \xi_{s} \cdot \partial) u_{s}^{\alpha}.
        \label{eq:uvar}
\end{equation}
It is easy to check that this variation preserves $u^{\alpha} 
u_{\alpha} = 1$.  

The variation of the proper density $N_{s}$ can be obtained by the 
requirement that the quantity
\begin{equation}
        N_{s} \left( \frac{dx_{s}}{d\tau} \cdot
                \frac{dx_{s}}{d\tau}\right)^{-1/2} d^{4} x
\end{equation}
should be kept intact (i.e., mass is conserved).  The factor in the bracket is 
the induced metric along the world line.  This requirement fixes the variation 
at a position following a fluid element along its worldline as $-N_{s} 
[(\partial \cdot \delta \xi_{s}) - u_{s\beta} (u_{s} \cdot \partial) \delta 
\xi_{s}^{\beta}] = -N_{s} \,[h_{s}^{\alpha\beta} \partial_{\alpha} \delta 
\xi_{s\beta}]$, and hence the Eulerian variation is given by
\begin{equation}
        \delta N_{s} = - (\delta \xi_{s} \cdot \partial) N_{s}
        - N_{s} \;{\sf h}_{s}:\partial\delta \xi_{s}.
        \label{eq:Nvar}
\end{equation}
It is straightforward to check that the above variations 
Eqs.~(\ref{eq:uvar}, \ref{eq:Nvar}) are consistent with the conservation 
law 
\begin{equation}
        \partial_{\alpha} J_{s}^{\alpha} = 0
        \label{eq:Jcon}
\end{equation}
of the flux four-vector $J_{s}^{\alpha} = N_{s} u_{s}^{\alpha}$.  It is useful 
to know its variation which can be easily calculated using 
Eqs.~(\ref{eq:uvar}, \ref{eq:Nvar}):
\begin{equation}
        \delta J_{s}^{\alpha} = 
        \partial_{\beta} (J_{s}^{\beta} \delta \xi_{s}^{\alpha} 
                -J_{s}^{\alpha} \delta \xi_{s}^{\beta}),
        \label{eq:Jvar}
\end{equation}
where the conservation law (\ref{eq:Jcon}) has been used.

Finally, the non-dissipative flow conserves entropy along the world 
line,
\begin{equation}
        (u_s \cdot \partial) S_s = 0.
        \label{eq:Scon}
\end{equation}
To be consistent with the variation Eq.~(\ref{eq:uvar}), we find
\begin{equation}
        \delta S_{s} = - (\delta \xi_{s} \cdot \partial) S_{s} .
        \label{eq:Svar}
\end{equation}
The expressions (\ref{eq:uvar}, \ref{eq:Nvar}, \ref{eq:Svar}) give the 
correct relativistic generalizations of the (non-relativistic) 
constrained Eulerian variations \cite{Newcomb}; see Appendix
\ref{sec:differential} for a geometric interpretation of 
Eqs.~(\ref{eq:uvar}, \ref{eq:Nvar}, \ref{eq:Jvar}, \ref{eq:Svar}).  
An alternative variational principle would introduce $\partial \cdot J_{s} = 0 
= u_{s} \cdot \partial S_{s}$ explicitly in the Lagrangian density by 
means of Lagrange multipliers \cite{Brown}.

\section{SELF-CONSISTENT NONLINEAR NEUTRINO-PLASMA FLUID EQUATIONS}
\label{sec:self}

We now proceed with the variational derivation of the dynamical 
equations for self-consistent neutrino-plasma fluid interactions. In deriving 
these equations, we use the thermodynamic relations (\ref{eq:first_law}
)-(\ref{eq:grad_mu}) as well as the continuity and entropy equations 
(\ref{eq:Jcon}, \ref{eq:Scon}) for each fluid species $s$. 

By re-arranging terms in the variational equation (\ref{eq:delta_l}) so as to 
isolate the variation four-vectors $\delta\xi_{s}$ and $\delta A$, we find
\begin{eqnarray}
\delta{\cal L} & \equiv & \partial\cdot{\cal J} \;-\; \sum_{s}\; \delta\xi_{s}
\cdot \left[\; \partial_{s}{\cal L} \;+\; \partial\cdot \left( u_{s}\;
\pd{{\cal L}}{u_{s}}\cdot {\sf h}_{s} \;-\; N_{s} \pd{{\cal L}}{N_{s}}\; 
{\sf h}_{s} \right) \;\right] \nonumber \\
 &   &\mbox{}+\; \delta A\cdot\left( \;\pd{{\cal L}}{A} \;-\; 2\;\partial\cdot
\pd{{\cal L}}{{\sf F}} \;\right),
\label{eq:new_delta_l}
\end{eqnarray}
where $\partial_{s}{\cal L} \equiv \partial N_{s}\,(\partial{\cal L}/\partial
N_{s}) + \partial u_{s}\cdot(\partial{\cal L}/\partial u_{s}) + \partial S_{s}
\,(\partial{\cal L}/\partial S_{s})$, and the Noether four-density ${\cal J}$ 
is expressed in terms of $\delta\xi_{s}$ and $\delta A$ as
\begin{equation}
{\cal J} \;\equiv\; \sum_{s}\; \left( u_{s}\;\pd{{\cal L}}{u_{s}}\cdot
{\sf h}_{s} \;-\; N_{s} \pd{{\cal L}}{N_{s}}\;{\sf h}_{s} \right)\cdot
\delta\xi_{s} \;+\; 2\,\pd{{\cal L}}{{\sf F}}\cdot \delta A.
\label{eq:Noether}
\end{equation}
When performing the variational principle (\ref{eq:vp}), with $\delta{\cal L}$ 
given by (\ref{eq:new_delta_l}), we only consider variations $\delta\xi_{s}$ 
and $\delta A$ which vanish on the integration boundary. Hence, the Noether 
density ${\cal J}$ in (\ref{eq:new_delta_l}) does not contribute to the 
dynamical equations.

\subsection{Plasma-Fluid Momentum Equation}

First, we derive the relativistic plasma-fluid four-momentum equation. Upon 
variation with respect to $\delta\xi_{\sigma}$ in (\ref{eq:vp}), we obtain
\begin{equation}
0 \;=\; \left( \partial N_{\sigma}\;\pd{{\cal L}}{N_{\sigma}} + 
\partial u_{\sigma}\cdot\pd{{\cal L}}{u_{\sigma}} +
\partial S_{\sigma}\;\pd{{\cal L}}{S_{\sigma}} \right) \;+\; \partial\cdot
\left( u_{\sigma}\;
\pd{{\cal L}}{u_{\sigma}}\cdot {\sf h}_{\sigma} \;-\; N_{\sigma} 
\pd{{\cal L}}{N_{\sigma}}\; {\sf h}_{\sigma} \right).
\label{eq:EL_sigma}
\end{equation}
Substitition of appropriate derivatives of the Lagrangian density ${\cal L}$ 
and using the constraint equations (\ref{eq:Jcon}, \ref{eq:Scon}) and
the thermodynamic 
relations (\ref{eq:first_law})-(\ref{eq:grad_mu}), this equation becomes the 
relativistic plasma-fluid four-momentum (covariant) equation
\begin{equation}
u_{\sigma}\cdot\partial \left( \mu_{\sigma}\;u_{\sigma} \right) \;=\; 
N_{\sigma}^{-1}\;\partial p_{\sigma} \;+\; \left(
q_{\sigma}\,{\sf F} \;+\; \sum_{\nu}\; G_{\sigma\nu}\,{\sf M}_{\nu} \right)
\cdot u_{\sigma},
\label{eq:cov_sigma}
\end{equation}
where 
\begin{equation}
M_{\nu}^{\alpha\beta} \;\equiv\; \partial^{\alpha}J_{\nu}^{\beta} \;-\;
\partial^{\beta}J_{\nu}^{\alpha}
\label{eq:tensor_nu}
\end{equation}
is an anti-symmetric tensor which represents the influence of the
neutrino background medium \cite{tensor}. This tensor satisfies the Maxwell-like equation
$\partial^{\rho}M_{\nu}^{\alpha\beta} + \partial^{\alpha} M_{\nu}^{\beta\rho} +
\partial^{\beta} M_{\nu}^{\rho\alpha} \equiv 0$ and its divergence is
$\partial_{\alpha}M_{\nu}^{\alpha\beta} \equiv \Box J_{\nu}^{\beta}$, 
where $\Box \equiv \partial\cdot\partial$ and the continuity equation $\partial
\cdot J_{\nu} = 0$ for the neutrino fluid was used. 

Separating the space and time components in (\ref{eq:cov_sigma}) (i.e., using
the $3 + 1$ notation), the spatial components of the plasma-fluid four-momentum 
equation (\ref{eq:cov_sigma}) yield
\begin{equation}
\left( \partial_{t} + {\bf v}_{\sigma}\bdot\nabla\right) \left(\, \mu_{\sigma}\,
\gamma_{\sigma}{\bf v}_{\sigma}/c^{2} \,\right) \;=\; -\,n_{\sigma}^{-1}\,
\nabla p_{\sigma} \;+\; q_{\sigma}\, \left( {\bf E} \;+\; 
\frac{{\bf v}_{\sigma}}{c} \btimes {\bf B} \right) \;+\; {\bf f}_{\sigma},
\label{eq:mom_eq_sigma}
\end{equation}
where ${\bf f}_{\sigma}$ is the neutrino-induced ponderomotive force (averaged
over neutrino species) on the plasma-fluid species $\sigma$, defined as 
\begin{equation}
{\bf f}_{\sigma} \;\equiv\; \sum_{\nu}\,G_{\sigma\nu} \left[\; -\,\left(
\nabla n_{\nu} \;+\; \frac{1}{c}\,\pd{\vb{J}_{\nu}}{t} \right) \;+\;
\frac{{\bf v}_{\sigma}}{c}\btimes \nabla\btimes\vb{J}_{\nu} \;\right].
\label{eq:force_sigma}
\end{equation}
The neutrino-induced ponderomotive force ${\bf f}_{\sigma}$ is composed of 
three terms: an electrostatic-like term $\nabla n_{\nu}$, an inductive-like 
term $\partial_{t} \vb{J}_{\nu}$, and a magnetic-like term $\nabla\btimes
\vb{J}_{\nu}$. This terminology is obviously motivated by the similarities with
the electromagnetic force on a charged particle.  In previous work by Silva 
{\it et al.} \cite{Silva}, only the electrostatic-like term is retained in the 
neutrino-induced ponderomotive force, i.e., the neutrino particle flux 
$\vb{J}_{\nu}$ is discarded under the assumption of isotropic neutrino and 
plasma fluids. 

\subsection{Neutrino-Fluid Momentum Equation}

Next, we derive the relativistic neutrino-fluid four-momentum equation; the
limiting case of zero neutrino masses is treated below (\ref{eq:force_nu}). 
Upon variation with respect to $\delta\xi_{\nu}$ in (\ref{eq:vp}), we obtain
\begin{equation}
0 \;=\; \left( \partial N_{\nu}\;\pd{{\cal L}}{N_{\nu}} + \partial u_{\nu}\cdot
\pd{{\cal L}}{u_{\nu}} + \partial S_{\nu}\;\pd{{\cal L}}{S_{\nu}} \right) \;+\;
\partial\cdot\left( u_{\nu}\;\pd{{\cal L}}{u_{\nu}}\cdot {\sf h}_{\nu} \;-\; 
N_{\nu} \pd{{\cal L}}{N_{\nu}}\; {\sf h}_{\nu} \right).
\label{eq:EL_nu}
\end{equation}
Substitution of derivatives of ${\cal L}$ and using the thermodynamic relations
(\ref{eq:first_law})-(\ref{eq:grad_mu}), this equation becomes the relativistic 
neutrino-fluid four-momentum equation
\begin{equation}
u_{\nu}\cdot\partial \left( \mu_{\nu}\;u_{\nu} \right) \;=\; N_{\nu}^{-1}\;
\partial p_{\nu} \;+\; \sum_{\sigma}\;G_{\sigma\nu}\;{\sf M}_{\sigma}\cdot 
u_{\nu},
\label{eq:cov_nu}
\end{equation}
where
\begin{equation}
M_{\sigma}^{\alpha\beta} \;\equiv\; \partial^{\alpha}J_{\sigma}^{\beta} 
\;-\; \partial^{\beta}J_{\sigma}^{\alpha}
\label{eq:tensor_sigma}
\end{equation}
is another anti-symmetric tensor which represents the influence of the
background medium. This tensor satisfies the Maxwell-like equation 
$\partial^{\rho}M_{\sigma}^{\alpha\beta} + \partial^{\alpha} 
M_{\sigma}^{\beta\rho} + \partial^{\beta} M_{\sigma}^{\rho\alpha} \equiv 0$ 
and its divergence is $\partial_{\alpha} M_{\sigma}^{\alpha\beta} \equiv \Box
J_{\sigma}^{\beta}$, where the continuity equation $\partial\cdot
J_{\sigma} = 0$ for the plasma fluid was used. In (\ref{eq:cov_nu}), we 
note that the neutrino fluid is thus under the influence of an 
electromagnetic-like force induced by nonuniform plasma flows. We also note 
that the symmetry between the ponderomotive forces (\ref{eq:tensor_nu}) and 
(\ref{eq:tensor_sigma}) is a result of the symmetry of the neutrino-plasma 
interaction term $(\sum_{\sigma} \sum_{\nu}\,G_{\sigma\nu} J_{\sigma}\cdot
J_{\nu})$ in the Lagrangian density (\ref{eq:lag_d}). 

Using the $3 + 1$ notation, the spatial components of neutrino-fluid 
four-momentum equation (\ref{eq:cov_nu}) yield
\begin{equation}
\left( \partial_{t} + {\bf v}_{\nu}\bdot\nabla \right) \left(\, \mu_{\nu}\, 
\gamma_{\nu}{\bf v}_{\nu}/c^{2} \,\right) \;=\; -\,n_{\nu}^{-1}\,\nabla p_{\nu} 
\;+\; {\bf f}_{\nu},
\label{eq:mom_eq_nu}
\end{equation}
where ${\bf f}_{\nu}$ is the plasma-induced ponderomotive force (averaged
over plasma-particle species) on the neutrino-fluid species $\nu$, defined as
\begin{equation}
{\bf f}_{\nu} \;\equiv\; \sum_{\sigma}\,G_{\sigma\nu} \left[\; -\, \left( 
\nabla n_{\sigma} \;+\; \frac{1}{c}\,\pd{\vb{J}_{\sigma}}{t} \right) \;+\; 
\frac{{\bf v}_{\nu}}{c}\btimes\nabla \btimes \vb{J}_{\sigma} \;\right].
\label{eq:force_nu}
\end{equation}
The plasma-induced ponderomotive force ${\bf f}_{\nu}$ on the neutrino fluid 
is composed of three terms: an electrostatic-like term $\nabla n_{\sigma}$, an 
inductive-like term $\partial_{t}\vb{J}_{\sigma}$, and a magnetic-like 
term $\nabla\btimes\vb{J}_{\sigma}$.

We now discuss the case of a cold ideal fluid composed of massless neutrinos.
Variation of the neutrino part of the Lagrangian density 
\[ {\cal L}_{\nu} \;\equiv\; -\,\frac{1}{2}\;n_{\nu}\lambda_{\nu}\, v_{\nu}
\cdot v_{\nu} \;-\; \sum_{\sigma}\; G_{\sigma\nu}\,J_{\sigma}\cdot J_{\nu} \]
with respect to $\delta\xi_{\nu}$ yields
\begin{equation}
\delta{\cal L}_{\nu} \;\equiv\; -\,n_{\nu}\lambda_{\nu}\;\delta v_{\nu}\cdot 
v_{\nu} \;-\; \sum_{\sigma}\,G_{\sigma\nu}\;J_{\sigma}\cdot\delta J_{\nu},
\label{eq:var_mld}
\end{equation}
where we used the constraint $v_{\nu}\cdot v_{\nu} \equiv 0$. Using 
$\delta J_{\nu} \equiv \partial\cdot (J_{\nu}\delta\xi_{\nu} -
\delta\xi_{\nu}J_{\nu})$ and $\delta v_{\nu}\cdot J_{\nu} =
\partial\cdot(J_{\nu}\delta\xi_{\nu})\cdot v_{\nu}$, the variation
equation (\ref{eq:var_mld}) becomes
\begin{equation}
\delta{\cal L}_{\nu} \;\equiv\; \partial\cdot{\cal J} \;+\; n_{\nu}\delta
\xi_{\nu} \cdot \left[\, v_{\nu}\cdot\partial(\lambda_{\nu}v_{\nu}) \;-\; 
\sum_{\sigma}\,G_{\sigma\nu}\; {\sf M}_{\sigma}\cdot v_{\nu} \,\right],
\label{eq:veq_massless}
\end{equation}
where the tensor ${\sf M}_{\sigma}$ is defined in (\ref{eq:tensor_sigma}) and
the Noether density is
\begin{equation}
{\cal J} \;\equiv\; \delta\xi_{\nu}\cdot \left[\, {\sf g}\; G_{\sigma\nu}
\,J_{\nu}\cdot J_{\sigma} \;-\; J_{\nu} \left( \lambda_{\nu}
v_{\nu} \;+\; \sum_{\sigma}\,G_{\sigma\nu}\,J_{\sigma} \right) \,\right].
\end{equation}
>From (\ref{eq:veq_massless}) the variational principle $\int \delta
{\cal L}_{\nu} \,d^{4}x = 0$ yields the cold neutrino fluid equation
\begin{equation}
v_{\nu}\cdot\partial(\lambda_{\nu}v_{\nu}) \;=\; \sum_{\sigma}\,G_{\sigma\nu}\;
{\sf M}_{\sigma}\cdot v_{\nu}.
\label{eq:nfe}
\end{equation}
In the cold-fluid limit, on the other hand, (\ref{eq:cov_nu}) yields $u_{\nu}
\cdot\partial(\gamma_{\nu}^{-1}\lambda_{\nu}\,u_{\nu}) \;=\; \sum_{\sigma}\,
G_{\sigma\nu}\; {\sf M}_{\sigma}\cdot u_{\nu}$, where $\lambda_{\nu}$ is the 
neutrino energy. By substituting $u_{\nu} \equiv \gamma_{\nu} v_{\nu}$ into 
this expression, we readily check that (\ref{eq:cov_nu}) and (\ref{eq:nfe})
are identical in the massless-neutrino  cold-fluid limit and that 
(\ref{eq:cov_nu}) can in fact be used to describe neutrino-fluid dynamics with 
arbitrary neutrino mass.

\subsection{Maxwell Equations}

The remaining equations are obtained from the variational principle 
(\ref{eq:vp}) upon variations with respect to the four-potential 
$\delta A^{\alpha}$. One thus obtains
\begin{equation}
0 \;=\; \pd{{\cal L}}{A} \;-\; 2\;\partial\cdot\pd{{\cal L}}{{\sf F}}.
\label{eq:EL_em}
\end{equation}
Substitution of derivatives of ${\cal L}$, this equation becomes the Maxwell 
equation
\begin{equation}
\partial\cdot {\sf F} \;=\; 4\pi \sum_{\sigma}\;q_{\sigma}\,J_{\sigma}.
\label{eq:cov_maxwell}
\end{equation}
Using the $3 + 1$ notation, we recover one half of the familiar Maxwell 
equations from (\ref{eq:cov_maxwell}). The other half is expressed in terms
of the Faraday tensor alone as 
\begin{equation}
\partial^{\rho}F^{\alpha\beta} + \partial^{\alpha}F^{\beta\rho} + 
\partial^{\beta}F^{\rho\alpha} \equiv 0,
\label{eq:Maxwell}
\end{equation}
which, using the $3 + 1$ notation, yields $\nabla\bdot{\bf B} = 0$ and
$\nabla\btimes{\bf E} + c^{-1}\partial_{t}{\bf B} = 0$.

\subsection{Energy-Momentum Conservation Laws}
\label{subsec:energy-momentum}

Since the dynamical equations (\ref{eq:EL_sigma}), (\ref{eq:EL_nu}) and 
(\ref{eq:EL_em}) are true for arbitrary variations $(\delta\xi_{\sigma}$,
$\delta\xi_{\nu})$ and $\delta A$ (subject to boundary conditions), the 
variational equation (\ref{eq:new_delta_l}) becomes
\begin{equation}
\delta{\cal L} \;\equiv\; \partial\cdot{\cal J},
\label{eq:Noether_eq}
\end{equation}
which we henceforth refer to as the Noether equation. We now discuss Noether 
symmetries of the Lagrangian density (\ref{eq:lag_d}) based on the Noether 
equation (\ref{eq:Noether_eq}).

For this purpose, we consider infinitesimal translations $x^{\alpha} 
\rightarrow x^{\alpha} + \delta x^{\alpha}$ generated by the infinitesimal 
displacement four-vector field $\delta x$. Under this transformation, the 
Lagrangian density ${\cal L}$ changes by
\begin{equation}
\delta{\cal L} \;\equiv\; -\,\partial\cdot(\delta x\;{\cal L}).
\label{eq:delta_l_delta_x}
\end{equation}
Next, we introduce the following explicit expressions for $(\delta\xi_{\sigma}, 
\delta\xi_{\nu})$ and $\delta A$ in terms of the infinitesimal generating 
four-vector $\delta x$:
\begin{equation}
\left. \begin{array}{rcl}
\delta\xi_{s} & \equiv & {\sf h}_{s}\cdot\delta x \\
 &  & \\
\delta A & \equiv & {\sf F}\cdot\delta x \;-\; \partial(A\cdot\delta x)
\end{array} \right\},
\label{eq:cv_delta_x}
\end{equation}
where the symmetric tensor ${\sf h}_{s}$ is defined in (\ref{eq:h}). (These
expressions are given geometric interpretations in Appendix 
\ref{sec:differential}.)

Substituting (\ref{eq:cv_delta_x}) in the Noether density (\ref{eq:Noether}), 
we find
\begin{equation}
{\cal J} \;=\; \left[ 2\,\pd{{\cal L}}{{\sf F}}\cdot{\sf F} \;+\; \sum_{s}\;
\left( u_{s}\;\pd{{\cal L}}{u_{s}}\cdot {\sf h}_{s} \;-\; N_{s}
\pd{{\cal L}}{N_{s}} \; {\sf h}_{s} \right) \right] \cdot\delta x \;+\; 2\,
\partial(A\cdot\delta x)\cdot \pd{{\cal L}}{{\sf F}},
\label{eq:init_Noether}
\end{equation}
where we have used the identity ${\sf h}_{s}\cdot{\sf h}_{s} = {\sf h}_{s}$ in
writing the second and third terms. Making use of the Maxwell equation 
(\ref{eq:cov_maxwell}), the last term in (\ref{eq:init_Noether}) can be 
re-arranged as 
\begin{equation}
2\,\partial(A\cdot\delta x)\cdot\pd{{\cal L}}{{\sf F}} \;=\; \partial\cdot 
\left[\; 2\,(A\cdot\delta x)\;\pd{{\cal L}}{{\sf F}} \;\right] \;-\; (A\cdot
\delta x)\;\pd{{\cal L}}{A}.
\label{eq:gauge}
\end{equation}
We now note that the expression for $\partial\cdot{\cal J}$ in 
(\ref{eq:Noether_eq}) is invariant under the transformation ${\cal J} 
\rightarrow {\cal J} + \partial\cdot {\sf K}$, where ${\sf K}$ is an 
antisymmetric tensor (for which $\partial_{\alpha\beta}^{2}\,K^{\alpha\beta} 
\equiv 0$) which vanishes on the integration boundary in (\ref{eq:vp}). Since 
the first term on the right side of (\ref{eq:gauge}) is such a term, it 
can be transformed away and the final expression for the Noether density is 
therefore
\begin{equation}
{\cal J} \;=\; \left[  2\,\pd{{\cal L}}{{\sf F}}\cdot {\sf F} \;-\;
\pd{{\cal L}}{A}\;A  \;+\; \sum_{s} \left( u_{s}\;\pd{{\cal L}}{u_{s}}\cdot
{\sf h}_{s} \;-\; N_{s} \pd{{\cal L}}{N_{s}}\; {\sf h}_{s} \right) \right]\cdot
\delta x.
\label{eq:final_Noether}
\end{equation}

Substituting (\ref{eq:delta_l_delta_x}) into (\ref{eq:Noether_eq}), the
Noether equation becomes $\partial\cdot({\cal J} + \delta x\,{\cal L}) = 0$. 
We define the symmetric energy-momentum tensor ${\sf T}$ from the expression
\begin{equation}
{\cal J} \;+\; \delta x\;{\cal L} \;\equiv\; -\,{\sf T}\cdot \delta x, 
\label{eq:def_T}
\end{equation}
where, using (\ref{eq:final_Noether}), the energy-momentum tensor ${\sf T}$ is 
explicitly given as
\begin{equation}
{\sf T} \;=\; -\,{\sf g}\;{\cal L} \;-\; \left( 2\,\pd{{\cal L}}{{\sf F}}\cdot
{\sf F} \;-\; \pd{{\cal L}}{A}\;A \right) \;-\; \sum_{s} \left(  u_{s}\;
\pd{{\cal L}}{u_{s}}\cdot{\sf h}_{s} \;-\; N_{s} \pd{{\cal L}}{N_{s}}\;
{\sf h}_{s} \right).
\label{eq:expl_T}
\end{equation}
For a constant translation $\delta x$, the Noether equation 
(\ref{eq:Noether_eq}) then becomes
\begin{equation}
0 \;=\; \partial\cdot {\sf T},
\label{eq:cons_T}
\end{equation}
where using the Lagrangian density (\ref{eq:lag_d}) and its derivatives in 
(\ref{eq:expl_T}), we find
\begin{eqnarray}
T^{\alpha\beta} & = & \frac{1}{4\pi} \left( F^{\alpha}_{\;\;\kappa}
F^{\kappa\beta} \;-\; \frac{g^{\alpha\beta}}{4}\;{\sf F}:{\sf F} \right) \;+\;
\sum_{s} \left( N_{s}\mu_{s} \; u_{s}^{\alpha}u_{s}^{\beta} \;-\; p_{s}\, 
g^{\alpha\beta} \right) \nonumber \\
 &   &\mbox{}+\; \sum_{\sigma}\sum_{\nu}\;G_{\sigma\nu}\; \left( 
J_{\sigma}^{\alpha}\, J_{\nu}^{\beta} \;+\; J_{\nu}^{\alpha}\,
J_{\sigma}^{\beta} \;-\; g^{\alpha\beta}\;J_{\sigma}\cdot J_{\nu} \right).
\label{eq:npe_T}
\end{eqnarray}
This energy-momentum tensor contains the usual terms associated with an 
electromagnetic field and a free relativistic ideal fluid 
\cite{Tolman,Weinberg,MTW}. It also contains the energy-momentum terms 
associated with collective neutrino-plasma interactions (third set of terms).

The energy-momentum transfer between the electromagnetic-plasma background
and the neutrinos can now be investigated. Such a process is relevant to 
supernova explosions, for example, where approximately 1$\%$ of the neutrino 
energy needs to be transferred to the surrounding plasma. First, we define the 
electromagnetic-plasma (EMP) energy-momentum tensor:
\begin{equation}
{\sf T}_{{\rm EMP}} \;\equiv\; \frac{1}{4\pi} \left( {\sf F}\cdot {\sf F} \;-\;
\frac{{\sf g}}{4}\;{\sf F}:{\sf F} \right) \;+\; \sum_{\sigma}\; \left( 
N_{\sigma}\mu_{\sigma} \; u_{\sigma}u_{\sigma} \;-\; p_{\sigma}\, {\sf g} 
\right) \;\equiv\; {\sf T}_{{\rm EM}} \;+\; {\sf T}_{{\rm P}},
\label{eq:EMP_T}
\end{equation}
and, using the exact energy-momentum conservation law (\ref{eq:cons_T}) as well 
as the dynamical equations (\ref{eq:cov_sigma}), (\ref{eq:cov_nu}) and 
(\ref{eq:cov_maxwell}), we find
\begin{equation}
\partial\cdot{\sf T}_{{\rm EMP}} \;=\; \sum_{\sigma} \left( \sum_{\nu}\; 
G_{\sigma\nu}\, {\sf M}_{\nu} \right) \cdot J_{\sigma}.
\label{eq:T_transfer}
\end{equation}
This equation describes how energy and momentum are transferred from the 
neutrinos to the electromagnetic field and the background plasma. Note how the 
transfer of energy-momentum between an electromagnetic-plasma and neutrinos is 
very much like the transfer of energy between a plasma (P) and an 
electromagnetic field (i.e., $\partial\cdot {\sf T}_{{\rm P}} = \sum_{\sigma}\, 
q_{\sigma} {\sf F}\cdot J_{\sigma}$) in the absence of neutrinos. 

We note that in addition to energy and momentum, wave action 
\cite{Brizard_Wurtele} can be transferred between the neutrinos and the 
electromagnetic-plasma background. In this case, electromagnetic waves and/or 
plasma waves can be excited by resonant three-wave processes.

\section{MAGNETIC FIELD GENERATION AND HELICITY PRODUCTION BY COLLECTIVE 
NEUTRINO-PLASMA INTERACTIONS}
\label{sec:magnetic}

An important application of the process of energy-momentum transfer associated 
with collective electromagnetic-plasma-neutrino interactions is the possibility 
of generating magnetic fields in an unmagnetized plasma as a result of 
collective neutrino-plasma interactions. Such process might be relevant to the 
problem of magnetogenesis and the production of magnetic helicity in the early 
universe \cite{string_MHD,helicity_1,helicity_2}. A similar process of 
magnetic-field generation has been observed in laser-plasma interactions 
\cite{GMA,Haines,MT,B_generation}.

According to our neutrino-plasma fluid model [based on (\ref{eq:cov_sigma}), 
(\ref{eq:cov_nu}, and (\ref{eq:cov_maxwell})], the strength of the magnetic 
field generated by neutrino-plasma interactions scales as the first power in 
the Fermi weak-interaction constant $G_{F}$. In what follows, we thus refer to 
magnetic fields generated by classical plasma processes (e.g., the 
Biermann-battery effect and the nonlinear dynamo effect) as zeroth-order fields 
while those generated by collective neutrino-plasma interactions as first-order 
fields.  Second-order fields, for example, might be produced by processes such 
as $\sigma^{\prime} \rightarrow \nu \rightarrow \sigma \rightarrow EM$, where 
the first plasma-particle species $(\sigma^{\prime})$ need not be charged 
(e.g., neutrons).

In this Section, we investigate the role played by collective neutrino-plasma
interactions in generating magnetic fields and magnetic helicity as well as
magnetic equilibrium.

\subsection{Magnetic-field Generation}

An equation describing magnetic-field generation resulting from collective 
neutrino-plasma interactions is derived as follows. We begin with Faraday's law
\begin{equation}
\pd{{\bf B}}{t} \;=\; -c\,\nabla\btimes{\bf E},
\label{eq:Farad_init}
\end{equation}
where for a given plasma-particle species $\sigma$ [using 
(\ref{eq:force_sigma})], the electric field ${\bf E}$ is expressed as
\begin{equation}
{\bf E} \;\equiv\; \frac{1}{q_{\sigma}} \left( {\bf F}_{\sigma} \;-\; 
{\bf f}_{\sigma} \right) \;-\; \frac{{\bf v}_{\sigma}}{c}\btimes{\bf B},
\label{eq:init_E}
\end{equation} 
where ${\bf f}_{\sigma}$ is the neutrino-induced ponderomotive force given by 
(\ref{eq:force_sigma}) and 
\begin{equation}
{\bf F}_{\sigma} \;\equiv\; \pd{{\bf P}_{\sigma}}{t} \;+\; {\bf v}_{\sigma}
\bdot\nabla {\bf P}_{\sigma} \;+\; n_{\sigma}^{-1}\,\nabla p_{\sigma}, 
\label{eq:F_sigma}
\end{equation}
with ${\bf P}_{\sigma} \equiv (\mu_{\sigma}/c^{2})\gamma_{\sigma}
{\bf v}_{\sigma}$ the generalized momentum for plasma-fluid species 
$\sigma$. 

Since the electric field ${\bf E}$ is common to all charged-particle species,
we multiply (\ref{eq:init_E}) on both sides by $q_{\sigma}^{2}$ and sum 
over all charged-particle species present in the plasma. Defining 
$\sum_{\sigma} q_{\sigma}^{2} \equiv Q^{2}$, the electric field ${\bf E}$ is 
then given as
\begin{equation}
{\bf E} \;=\; \sum_{\sigma}\; \frac{q_{\sigma}}{Q^{2}}\; \left( 
{\bf F}_{\sigma} \;-\; {\bf f}_{\sigma} \right) \;-\; \left( \sum_{\sigma}\; 
\frac{q_{\sigma}^{2}{\bf v}_{\sigma}}{cQ^{2}} \right) \btimes {\bf B}.
\end{equation}
Substituting explicit expressions for ${\bf F}_{\sigma}$ and ${\bf f}_{\sigma}$,
we obtain
\begin{equation}
{\bf E} \;\equiv\; \left. \left. \sum_{\sigma}\; \frac{q_{\sigma}}{Q^{2}} 
\right[ \partial_{t} \vb{\Pi}_{\sigma} - {\bf v}_{\sigma}\btimes\nabla\btimes
\vb{\Pi}_{\sigma} + \nabla\chi_{\sigma} + S_{\sigma}\,\nabla(
\gamma_{\sigma}^{-1}T_{\sigma}) \right] \;-\; \left( \sum_{\sigma}\;
\frac{q_{\sigma}^{2} {\bf v}_{\sigma}}{cQ^{2}} \right)\btimes {\bf B},
\label{eq:final_E}
\end{equation}
where $\gamma_{\sigma}^{-1}T_{\sigma}$ is the temperature in the lab reference
frame and
\begin{equation}
\left. \begin{array}{rcl}
\vb{\Pi}_{\sigma} & \equiv & {\bf P}_{\sigma} \;+\; \sum_{\nu}\,G_{\sigma\nu}
\;\vb{J}_{\nu}/c \\
 &   & \\
\chi_{\sigma} & \equiv & \sum_{\nu}\; G_{\sigma\nu}\,n_{\nu} \;+\; 
\gamma_{\sigma}\,\mu_{\sigma} \;-\; \gamma_{\sigma}^{-1}T_{\sigma}S_{\sigma}
\end{array} \right\}.
\label{eq:pichi}
\end{equation}
Eq.~(\ref{eq:final_E}) can then be substituted for the electric field into 
Faraday's law (\ref{eq:Farad_init}) to give
\begin{eqnarray}
\pd{{\bf B}}{t} & = &  \sum_{\sigma} \frac{cq_{\sigma}}{Q^{2}} \left[\;
\nabla\left(\gamma_{\sigma}^{-1}T_{\sigma}\right) \btimes \nabla S_{\sigma}
\;-\;  \nabla\btimes \left( \partial_{t}{\bf P}_{\sigma} \;-\; {\bf v}_{\sigma}
\btimes\nabla\btimes{\bf P}_{\sigma} \right) \;\right] \nonumber \\
 &   &\mbox{}+\;\sum_{\sigma} \frac{q_{\sigma}^{2}}{Q^{2}} \;\nabla
\btimes \left( {\bf v}_{\sigma}\btimes {\bf B} \right) \;-\; \sum_{\nu}
\sum_{\sigma}\; \frac{q_{\sigma}G_{\sigma\nu}}{Q^{2}}\; \nabla\btimes \left( 
\partial_{t}\vb{J}_{\nu} \;-\; {\bf v}_{\sigma}\btimes\nabla\btimes
\vb{J}_{\nu} \right).
\label{eq:Farad_final}
\end{eqnarray}
The first collection of terms (linear in $q_{\sigma}$) on the right side of 
(\ref{eq:Farad_final}) includes the so-called Biermann-battery term ($\nabla 
n_{\sigma}^{-1}\btimes\nabla T_{\sigma}$) \cite{Haines,MT,Kulsrud} while the 
second term (proportional to $q_{\sigma}^{2}$) represents the nonlinear dynamo 
effect. These classical (zeroth-order) terms have been known to play important 
roles in the generation of magnetic fields during laser-plasma interactions 
\cite{GMA,Haines,MT,B_generation} as well as the evolution of cosmic and 
galactic magnetic fields \cite{Kulsrud}.

The last collection of terms (proportional to $q_{\sigma}G_{\sigma\nu}$) in 
(\ref{eq:Farad_final}) are associated with collective neutrino-plasma 
interactions and are completely new. Here, the neutrino-flux vorticity 
($\nabla\btimes \vb{J}_{\nu}$) plays a fundamental role in generating 
first-order magnetic fields; such terms are completely missing from previous 
works \cite{Shukla_97,Shukla_98}. 

According to (\ref{eq:Farad_final}), the electrostatic part of the 
neutrino-induced ponderomotive force (\ref{eq:force_sigma}) does not play any 
role in generating magnetic fields. Indeed, for each neutrino-fluid species
$\nu$, we have $\nabla\btimes [(\sum_{s}\,q_{s}G_{s\nu})\,\nabla n_{\nu}] = 0$,
independent of the plasma-fluid composition. The neutrino-induced ponderomotive 
force on plasma particles of species $\sigma$ actually given in 
\cite{Bingham,Shukla_97,Shukla_98} is $-n_{\sigma}^{-1}\,(\sum_{s^{\prime}}
G_{s^{\prime}\nu}\, n_{s^{\prime}})\,\nabla n_{\nu} \equiv 
{\bf f}_{\sigma}^{(B)}$; this expression improperly involves a sum of 
plasma-particle species $(\sum_{s^{\prime}})$ instead of the sum over neutrino 
species $(\sum_{\nu})$ as it appears in (\ref{eq:force_sigma}). Shukla 
{\it et al.} \cite{Shukla_98} then go on to develop a model for magnetic-field 
generation based on the fact that $\nabla\btimes {\bf f}_{\sigma}^{(B)} \neq 0$ 
for a plasma with multiple particles species. Since the sum over 
plasma-particle species ($\sum_{s^{\prime}}$) appearing in 
${\bf f}_{\sigma}^{(B)}$ is inappropriate, however, the conclusion drawn by 
Shukla {\it et al.} \cite{Shukla_98} that magnetic fields can be generated in a 
plasma composed of neutrons ($\sigma = n$) and electrons ($\sigma = e$) by 
terms such as $\nabla(n_{n}/n_{e}) \btimes\nabla n_{\nu}$ is incorrect 
\cite{neutron}. 

For a primordial plasma, we note that the Biermann-battery term could be small
unless the terms $\nabla(\gamma_{\sigma}^{-1}T_{\sigma}) \btimes \nabla 
S_{\sigma}$ and $\nabla(\gamma_{\ov{\sigma}}^{-1}T_{\ov{\sigma}}) \btimes 
\nabla S_{\ov{\sigma}}$ are in opposite directions whereas the nonlinear
dynamo requires net plasma flow. Using the identities (\ref{eq:identity_PG}), 
on the other hand, we note that particles $(\sigma)$ and anti-particles 
$(\ov{\sigma})$ of the same family $(\sigma,\ov{\sigma})$ contribute equally to 
the generation of first-order magnetic fields in a primordial plasma since 
\begin{equation}
\left. \begin{array}{rcl}
\sum_{s = \sigma,\ov{\sigma}}\; q_{s}\,G_{s\nu} & = & 2\, q_{\sigma}\,
G_{\sigma\nu} \\
 &   & \\
\sum_{s = \sigma,\ov{\sigma}}\; q_{s}\,G_{s\nu}{\bf v}_{s} & = & q_{\sigma}\,
G_{\sigma\nu}\,({\bf v}_{\sigma} + {\bf v}_{\ov{\sigma}})
\end{array} \right\}.
\label{eq:qG}
\end{equation}
This remark is especially relevant to the problem of magnetogenesis in the
early universe. Conversely, we note from (\ref{eq:Farad_final}) that a
time-dependent magnetic field {\it automatically} generates neutrino-flux
vorticity $\nabla \btimes\vb{J}_{\nu}$. Hence, the usual assumption that the
neutrino distribution is isotropic \cite{Silva} appears to be inconsistent with
first-order magnetic-field generation by first-order collective neutrino-plasma
interactions.

\subsection{Magnetic Helicity Production}

Another quantity intimately associated with magnetic-field generation is the
generation of magnetic helicity
\begin{equation}
H \;\equiv\; \int_{V} {\bf A}\bdot {\bf B}\; d^{3}x,
\label{eq:helicity}
\end{equation}
where $V$ is the three-dimensional volume which encloses the magnetic field
lines; to ensure that this definition of magnetic helicity be gauge invariant,
we require that ${\bf B}\cdot\wh{n} = 0$, where $\wh{n}$ is a unit vector 
normal to the surface $\partial V$. Magnetic helicity is a measure of 
knottedness (or flux linkage) in the magnetic field \cite{Moffatt_Ricca};
hence a uniform magnetic field (or more generally a magnetic field which has a 
global representation in terms of Euler potentials $\alpha$ and $\beta$ as 
${\bf B} \equiv \nabla\alpha\btimes \nabla\beta$) has zero helicity. The 
production of magnetic helicity is therefore an indication that the spatial 
structure (and topology) of the magnetic field is becoming more complex. It is
expected that this feature in turn plays a fundamental role in the formation
of large-scale structure in the universe \cite{Peebles}. 

The time evolution of the magnetic helicity (\ref{eq:helicity}) leads to the 
equation
\begin{equation}
\frac{dH}{dt} \;=\; -2c\,\int_{V}\; {\bf E}\bdot{\bf B}\, d^{3}x \;-\; c\,
\int_{\partial V}\; \left( \phi\,{\bf B} \;+\; {\bf E}\btimes{\bf A} \right)
\bdot \wh{n}\,d^{2}x,
\label{eq:hel_eq}
\end{equation}
where integration by parts was performed in obtaining the surface term. Taking 
the integration volume $V$ arbitrarily large (or requiring that ${\bf E}$ be
parallel to $\wh{n}$ in addition to ${\bf B}\bdot\wh{n} = 0$), we find that the 
surface term vanishes and we are left only with the first term in 
(\ref{eq:hel_eq}). If we now substitute (\ref{eq:final_E}) into 
(\ref{eq:hel_eq}), we obtain 
\begin{equation}
\frac{dH}{dt} \;=\; -\, \sum_{\sigma} \frac{2q_{\sigma}c}{Q^{2}}\; \left. \left.
\int_{V}\; {\bf B}\bdot\right[\, \partial_{t}\vb{\Pi}_{\sigma} - 
{\bf v}_{\sigma}\btimes \nabla\btimes\vb{\Pi}_{\sigma} + \nabla\chi_{\sigma} + 
S_{\sigma}\,\nabla (\gamma_{\sigma}^{-1}T_{\sigma}) \,\right] d^{3}x,
\label{eq:H_eq}
\end{equation}
where $\vb{\Pi}_{\sigma}$ and $\chi_{\sigma}$ are defined in (\ref{eq:pichi}).
Since the term ${\bf B}\bdot\nabla\chi_{\sigma}$ can be written as an exact
divergence, it does not contribute to the production of magnetic helicity.
Furthermore, since temperature gradients along the magnetic field, ${\bf B}\bdot
\nabla(\gamma_{\sigma}^{-1}T_{\sigma})$, vanish in the absence of dissipative
effects the last term in (\ref{eq:H_eq}) drops out. Hence, magnetic helicity 
production is governed by the equation
\begin{equation}
\frac{dH}{dt} \;=\; -\, \sum_{\sigma} \frac{2q_{\sigma}c}{Q^{2}}\; \int_{V}\;
{\bf B}\bdot\left(\, \partial_{t}\vb{\Pi}_{\sigma} - {\bf v}_{\sigma}\btimes
\nabla\btimes\vb{\Pi}_{\sigma} \right).
\label{eq:MHP}
\end{equation}
This equation states that helicity production can occur in the presence of
(zeroth-order) non-trivial flows \cite{Moffatt_Ricca} and/or (first-order)
nonuniform neutrino flux. 

It has been pointed out that magnetic helicity plays an important role in 
allowing energy to be transferred from small to large scales by a process 
called inverse cascade. Thus neutrino-flux vorticity leads to the generation of 
small-scale magnetic fields, first, and then to the production of magnetic 
helicity. The production of magnetic helicity, on the other hand, converts the 
small-scale magnetic fields to large-scale magnetic fields which are expected 
to play a fundamental role in the problem of structure formation in the 
early universe.  The magnetic helicity production described by (\ref{eq:MHP}) 
involves a multi-species fluid picture. A more standard description is based 
on the magnetohydrodynamic (MHD) equations in which plasma flows are averaged 
over particle species. Future work will proceed by deriving ideal neutrino-MHD 
equations.

\subsection{Magnetic Equilibrium in a Magnetized Plasma and Neutrino Fluid}

When gravitational effects can be ignored, plasmas can be confined by magnetic
fields. Such an equilibrium is established by balancing the (outward) kinetic 
pressure gradient with the (inward) magnetic pressure gradient. We now
investigate how magnetic equilibria are modified by the presence of
neutrino fluxes.

The equation for magnetic equilibrium involving magnetic fields associated with 
neutrino-plasma interactions can be obtained by multiplying (\ref{eq:init_E}) 
with $q_{\sigma}n_{\sigma}$ and summing over the charged-particle species only. 
In a time-independent equilibrium $(\partial/\partial t \equiv 0)$ involving a 
quasi-neutral plasma (where $\sum_{\sigma}\, q_{\sigma}n_{\sigma} = 0$), a 
static magnetic field ${\bf B}$ and time-independent neutrino fluids, we find 
the following equilibrium condition
\begin{eqnarray}
\frac{{\bf J}}{c}\btimes{\bf B} & = & \nabla\bdot \left[\; \sum_{\sigma}
\left( n_{\sigma}{\bf v}_{\sigma}{\bf P}_{\sigma} + {\bf I}\,p_{\sigma} \right) 
\;\right] \;+\; \sum_{\nu} \left[\; \left( \sum_{\sigma}\,n_{\sigma}\,
G_{\sigma\nu} \right) \nabla n_{\nu} \;\right] \nonumber \\
 &   &\mbox{}-\;  \sum_{\nu} \left[\; \left( \sum_{\sigma}\, G_{\sigma\nu}\;
\frac{n_{\sigma}{\bf v}_{\sigma}}{c} \right) \btimes\nabla\times 
\vb{J}_{\nu} \;\right],
\label{eq:equilibrium}
\end{eqnarray}
where ${\bf J} \equiv (c/4\pi)\,\nabla\btimes{\bf B} = \sum_{\sigma} q_{\sigma}
\vb{J}_{\sigma}$ is the current density flowing in a time-independent 
magnetized plasma. The first term on the right side of (\ref{eq:equilibrium}) 
represents the classical term associated with equilibrium in a magnetized 
plasma. The second and third terms denote first-order neutrino-plasma 
contributions to magnetic-field equilibrium.

Shukla {\it et al.} \cite{Shukla_97} derived a similar equilibrium condition 
with only the electrostatic-like term present on the right side 
(\ref{eq:equilibrium}). For a primordial plasma, using (\ref{eq:identity_PG}),
we note that the neutrino-induced electrostatic-like term once again vanishes 
from the magnetic-field generation picture. Hence, whereas the second term in 
(\ref{eq:equilibrium}) vanishes for a primordial plasma, the third term on the 
right side of (\ref{eq:equilibrium}), however, does not.  Magnetic equilibrium 
in a primordial neutrino-plasma is thus described by the balance equation
\begin{equation}
\sum_{\sigma}\; \frac{\vb{J}_{\sigma}}{c}\btimes \left( q_{\sigma}\,{\bf B} 
\;+\; \sum_{\nu}\;G_{\sigma\nu}\;\nabla\btimes\vb{J}_{\nu} \right) \;=\; 
\nabla\bdot \left[\; \sum_{s = \sigma,\ov{\sigma}}\left( n_{s}{\bf v}_{s}
{\bf P}_{s} + {\bf I}\,p_{s} \right) \;\right],
\label{eq:equilibrium_prim}
\end{equation}
where summation over species on the left side of (\ref{eq:equilibrium_prim}) 
involves only particle species, while the summation on the right side involves 
particle and anti-particle species. Once again, neutrino-flux vorticity 
$\nabla\btimes\vb{J}_{\nu}$ plays a fundamental role in collective 
neutrino-plasma interactions in the presence of an electromagnetic field.

\section{SUMMARY AND FUTURE WORK}
\label{sec:summary}

We now summarize our work and discuss future work. The model for collective 
neutrino-plasma interactions presented in this work is based on the nonlinear 
dissipationless fluid equations (\ref{eq:cov_sigma}), (\ref{eq:cov_nu}) and 
(\ref{eq:cov_maxwell}). These equations are derived from a variational 
principle based on the relativistic covariant Lagrangian density 
(\ref{eq:lag_d}). An exact energy-momentum conservation law (\ref{eq:cons_T})
is obtained by Noether method with the energy-momentum tensor for
self-consistent collective neutrino-plasma interactions in the presence of an
electromagnetic field is given by (\ref{eq:npe_T}). New ponderomotive forces 
acting on the plasma-neutrino fluids, which are absent from previous works 
\cite{Bingham,Shukla_97,Shukla_98}, are given by (\ref{eq:tensor_nu}) and 
(\ref{eq:tensor_sigma}) [or (\ref{eq:force_sigma}) and (\ref{eq:force_nu}), 
respectively]. In Eqs.~(\ref{eq:Farad_final}) and (\ref{eq:equilibrium}), we 
have demonstrated the crucial role played by neutrino-flux vorticity 
$(\nabla\btimes \vb{J}_{\nu})$ in the processes of magnetic-field 
generation and magnetic-helicity production in neutrino-plasma fluids. 

In future work, we plan to further investigate the importance of the new
neutrino-induced ponderomotive terms associated with neutrino fluxes. For this 
purpose, it might also be useful to derive from ideal 
neutrino-magnetohydrodynamic equations from (\ref{eq:cov_sigma}), 
(\ref{eq:cov_nu}) and (\ref{eq:cov_maxwell}). Using the new mechanisms for 
magnetic-field generation and magnetic-helicity production proposed in 
(\ref{eq:Farad_final}) and (\ref{eq:MHP}), respectively, we plan to investigate 
the problem of magnetogenesis in the early universe. As another application, we 
plan to investigate neutrino-plasma three-wave interactions leading to the 
excitation of various plasma waves in unmagnetized and magnetized plasmas; such 
transfer processes could be important during supernova explosions.

\acknowledgments
%\section*{ACKNOWLEDGMENTS}

This work was performed under Department of Energy Contracts 
No.~PDDEFG-03-95ER-40936 and DE-AC03-76SF00098, in part by the 
National Science Foundation under grant PHY-95-14797, and in part also 
by Alfred P. Sloan Foundation.

\appendix

\section{DIFFERENTIAL GEOMETRIC FORMULATION OF CONSTRAINED VARIATIONS}
\label{sec:differential}

In this Appendix, the geometric interpretation of the constrained 
variations (\ref{eq:uvar}, \ref{eq:Nvar}, \ref{eq:Jvar}, \ref{eq:Svar}) is 
given in terms of Lie derivatives along the virtual displacement 
four-vector $\delta\xi$.  Since the variation of a fluid field is only its 
infinitesimal displacement, all covariant quantities 
are varied by their Lie derivatives with respect to the virtual 
displacement four-vector $\delta\xi$.  Here, we use the following 
definition of the Lie derivative on the $k$-form $\alpha$ along the 
four-vector $\delta\xi$, denoted ${\bf L}_{\delta\xi}\alpha$ 
\cite{AMR}:
\begin{equation}
{\bf L}_{\delta\xi}\alpha \;\equiv\; {\bf i}_{\delta\xi}\cdot d\alpha \;+\; d
\left( {\bf i}_{\delta\xi}\cdot\alpha \right).
\label{eq:def_lie}
\end{equation}
Here, $d\alpha$ is a $(k+1)$-form while ${\bf i}_{\delta\xi}\cdot\alpha$ is a
$(k-1)$-form representing the contraction of the four-vector $\delta\xi$ with
the $k$-form $\alpha$. By definition, if $\alpha = \varphi$ is a scalar field
(i.e., a zero-form), ${\bf i}_{\delta\xi}\cdot\varphi \equiv 0$.

The constrained variation $\delta S = -\,\delta\xi\cdot\partial S$ for the
entropy $S$ [(\ref{eq:Svar})] is consistent with its geometric interpretation 
as a scalar field:
\begin{equation}
\delta S \;\equiv\; -\,{\bf L}_{\delta\xi}S \;=\; -\,\delta\xi\cdot\partial S,
\label{eq:lie_S}
\end{equation}
where ${\bf i}_{\delta\xi}\cdot S \equiv 0$ and ${\bf i}_{\delta\xi}\cdot dS 
\equiv (\delta\xi\cdot\partial) S$.

The geometric interpretation of the particle flux $J^{\alpha} \equiv N 
u^{\alpha}$ is given as the components of the three-form 
$J = (1/3!) \epsilon_{\alpha\beta\kappa\lambda}J^\alpha dx^\beta
dx^\kappa dx^\lambda$. The constrained variation of the particle-flux 
four-vector is defined as
\begin{equation}
\delta J \;\equiv\; -\,{\bf L}_{\delta\xi} J.
\label{eq:lie_vector}
\end{equation}
Since $dJ \equiv (\partial\cdot J)\,\vb{\Omega}$ with the 
volume four-form $\vb{\Omega} \equiv dx^{0}\wedge dx^{1}\wedge dx^{2}\wedge
dx^{3}$, and hence $dJ=0$ due to the continuity equation,
we obtain $\delta J = - d({\bf i}_{\delta\xi} \cdot J)$, or
\begin{equation}
        \delta J^{\alpha} = 
        \partial_{\beta} (J^{\beta} \delta \xi^{\alpha} 
                -J^{\alpha} \delta \xi^{\beta}),
\label{eq:lie_gamma}
\end{equation}
which is Eq.~(\ref{eq:Jvar}) itself.
>From this variation, one can easily compute the variations of $N = 
\sqrt{J^{\alpha} J_{\alpha}}$ and $u^{\alpha} = J^{\alpha}/N$ leading
(\ref{eq:Nvar}) and (\ref{eq:uvar}), respectively.

In Sec.~\ref{subsec:energy-momentum}, we consider infinitesimal 
translations $x^{\alpha} \rightarrow x^{\alpha} + \delta x^{\alpha}$ 
generated by the infinitesimal displacement four-vector $\delta x$.  
Under this transformation, the Lagrangian density ${\cal L}$ changes 
by $\delta{\cal L} \equiv -\,\partial\cdot(\delta x \;{\cal L})$.  
This expression is consistent with the geometric interpretation of 
${\cal L}$ as a density in four-dimensional space, i.e.,
\begin{equation}
\delta{\cal L}\,\vb{\Omega} \;\equiv\; -\,{\bf L}_{\delta x}({\cal L}\,
\vb{\Omega}), 
\label{eq:lie_lag}
\end{equation}
where ${\bf L}_{\delta x}$ is the Lie derivative with respect to $\delta x$. 
Here, using ${\bf i}_{\delta x}\cdot d({\cal L}\,\vb{\Omega}) = 0$ and 
\begin{equation}
d\left[ {\bf i}_{\delta x}\cdot \left( {\cal L}\,\vb{\Omega}\right) \right] 
\;=\;
d \left( {\cal L}\;\delta x\cdot\omega \right) \;\equiv\; \partial\cdot(\delta 
x\; {\cal L})\;\vb{\Omega},
\end{equation}
we easily recover (\ref{eq:delta_l_delta_x}).

Next, the expressions for $\delta A$ is given in (\ref{eq:cv_delta_x}).  Here, 
the electromagnetic four-potential $A$ appears as the the components of the 
one-form $A\cdot dx$. Thus 
\begin{equation}
\delta A\cdot dx \;\equiv\; -{\bf L}_{\delta x}(A\cdot dx).
\label{eq:lie_A}
\end{equation}
Since ${\bf i}_{\delta x}\cdot d(A\cdot dx) = -\,({\sf F}\cdot\delta x)\cdot
dx$ and $d[{\bf i}_{\delta x}\cdot(A\cdot dx)] = d(A\cdot dx)$, we easily
recover (\ref{eq:cv_delta_x}) for the four-potential $A$. We note that the
expression $\delta\xi \equiv {\sf h}\cdot\delta x$ given in 
(\ref{eq:cv_delta_x}) is consistent with the expressions $\delta S = -\,
{\bf L}_{\delta\xi}S \equiv -\,{\bf L}_{\delta x}S$ and $\delta J\cdot
\omega = -\,{\bf L}_{\delta\xi}(J\cdot\omega) \equiv -\,{\bf L}_{\delta x}(
J\cdot\omega)$. 

\vfill\eject


\begin{thebibliography}{99}

\bibitem{KT} E.W.~Kolb and M.S.~Turner, {\it The Early Universe} 
(Addison-Wesley, Redwood City, California, 1990).

\bibitem{Peebles} P.J.E.~Peebles, {\it Principles of Physical Cosmology} 
(Princeton University Press, Princeton, New Jersey, 1993).

\bibitem{primordial} A primordial plasma is defined here as a quasi-neutral 
plasma composed of particles and anti-particles of the same family.

\bibitem{ST} S.L.~Shapiro and S.A.~Teukolsky, {\it Black Holes, White Dwarfs,
and Neutron Stars} (Wiley, New York, 1983), chap.~18.

\bibitem{CHB} J.~Cooperstein, L.J.~van der Horn and E.A.~Baron, Ap.~J {\bf 309},
653 (1986).

\bibitem{Cooperstein} J.~Copperstein, Phys.~Rep.~{\bf 163}, 95 (1988).

\bibitem{Brizard} A.J.~Brizard, Phys.~Plasmas {\bf 5}, 1110 (1998).

\bibitem{photon_neutrino} D.A.~Dicus and W.W.~Repko, Phys.~Rev.~Lett.~{\bf 79},
569 (1997); D.~Seckel, Phys.~Rev.~Lett. {\bf 80}, 900 (1998); R.~Shaisultanov, 
{\it ibid}, 1586 (1998).

\bibitem{Taylor} J.C.~Taylor, {\it Gauge theories of weak interactions}
(Cambridge Univ.~Press, Cambridge, 1976), chap.~8. 

\bibitem{electroweak} S.~Weinberg, {\it The Quantum Theory of Fields, vol.~II} 
(Cambridge University Press, Cambridge, 1996), sec.~21.3.

\bibitem{PP} P.B.~Pal and T.N.~Pham, Phys.~Rev.~D {\bf 40}, 259 (1989).

\bibitem{NR} D.~Notzold and G.~Raffelt, Nucl.~Phys.~B {\bf 307}, 924 (1988).

\bibitem{EC} S.~Esposito and G.~Capone, Z.~Phys.~C {\bf 70}, 55 (1996).

\bibitem{Bingham} R.~Bingham, H.A.~Bethe, J.M.~Dawson, P.K.~Shukla and J.J.~Su,
Phys.~Lett. A {\bf 220}, 107 (1996).

\bibitem{Brizard_Wurtele} A.J.~Brizard and J.S.~Wurtele, Phys. Plasmas {\bf 6},
1323 (1999).

\bibitem{Tajima} C.H.~Lai and T.~Tajima (unpublished work, 1994). This work
appears in T.~Tajima and K.~Shibata, {\it Plasma Astrophysics} (Addison-Wesley,
Reading, MA, 1997), sec.~5.4.

\bibitem{Silva} L.O.~Silva, R.~Bingham, J.M.~Dawson and W.B.~Mori, Phys.~Rev. E
{\bf 59}, 2273 (1999).

\bibitem{Silva_prl} L.O.~Silva, R.~Bingham, J.M.~Dawson, J.T.~Mendonca, and
P.K.~Shukla, Phys.~Rev.~Lett. {\bf 83}, 2703 (1999).

\bibitem{potential} See, {\it e.g.}, H.~Nunokawa, V.B.~Semikoz, A.Y.~Smirnov 
and J.W.F.~Valle, Nucl.~Phys. B {\bf 501}, 17 (1997); J.M.~Laming, 
Phys.~Lett.~A {\bf 255}, 318 (1999).

\bibitem{MSW} L.~Wolfenstein, Phys.~Rev.~D {\bf 17}, 2369 (1978).

\bibitem{Bethe} H.A.~Bethe, Phys.~Rev.~Lett. {\bf 56}, 1305 (1986).

\bibitem{mode_conv} R.~Bingham, R.A.~Cairns, J.M.~Dawson, R.O.~Dendy, 
C.N.~Lashmore-Davies and V.N.~Tsytovich, Phys.~Lett. A {\bf 232}, 257 (1997).

\bibitem{Balantekin} For a review, see A.B.~Balantekin, Phys.~Rep.~{\bf 315}, 
123 (1999)

\bibitem{hot} We note that the relativistic correction $\vb{J}_{\sigma}
\bdot{\bf v}_{\nu}/c$ scales as $\beta_{\nu}\beta_{\sigma}$ relative to the 
density term $n_{\sigma}$. Higher-order terms not shown in (\ref{eq:pot_nu})
\cite{NR} involve terms which scale as $E_{\nu}E_{\sigma}/m_{W}^{2}c^{4}$, 
where $m_{W}c^{2}$ ($\approx 80$ GeV) is the rest energy of the $W$ 
boson and $E_{s}$ is the typical particle energy for species $s$. Note that 
since $G_{F} \propto m_{W}^{-2}$ \cite{Taylor}, the higher-order corrections 
can also be called second-order corrections. Since $\beta_{\nu}$ is expected to 
be close to unity, we find that the relativistic correction kept in 
(\ref{eq:pot_nu}) is dominant over the second-order correction provided 
$\beta_{\sigma} \gg E_{\nu} E_{\sigma}/m_{W}^{2}c^{4}$; for neutrino and plasma 
characteristic energies less than 100 MeV, this condition is well satisfied if
$\beta_{\nu}\beta_{\sigma} > 10^{-4}$.

\bibitem{magnetic} P.~Meszaros, {\it High-Energy Radiation From Magnetized 
Neutron Stars} (University of Chicago Press, Chicago, 1992), chap.~2.

\bibitem{GMA} L.~Gorbunov, P.~Mora and T.M.~Antonsen, Phys.~Rev.~Lett.
{\bf 76}, 2495 (1996); Phys.~Plasmas {\bf 4}, 4358 (1997).

\bibitem{Haines} M.G.~Haines, Phys.~Rev.~Lett.~{\bf 78}, 254 (1997).

\bibitem{MT} R.J.~Mason and M.~Tabak, Phys.~Rev.~Lett.~{\bf 80}, 524 (1998).

\bibitem{B_generation} M.~Borghesi, A.J.~Mackinnon, R.~Gaillard, O.~Willi,
A.~Pukhov and J.~Meyer-ter-Vehn, Phys.~Rev.~Lett.~{\bf 80}, 5137 (1998).

\bibitem{Shukla_97} P.K.~Shukla, L.~Stenflo, R.~Bingham, H.A.~Bethe, 
J.M.~Dawson and J.T.~Mendon\c{c}a, Phys.~Lett. A {\bf 233}, 181 (1997).

\bibitem{Shukla_98} P.K.~Shukla, R.~Bingham, J.T.~Mendon\c{c}a and L.~Stenflo,
Phys.~Plasmas {\bf 5}, 2815 (1998).

\bibitem{Brown} J.D.~Brown, Class.~Quantum Grav.~{\bf 10}, 1579 (1993).

\bibitem{Achterberg} A.~Achterberg, Phys.~Rev. A {\bf 28}, 2449 (1983).

\bibitem{Whitham} G.B.~Whitham, {\it Linear and Nonlinear Waves} (Wiley,
New York, 1974), sec.~11.7.

\bibitem{Similon} P.L.~Similon, Phys.~Lett.~A {\bf 112}, 33 (1985).

\bibitem{Mills} R.~Mills, Am.~J.~Phys.~{\bf 57}, 493 (1989).

\bibitem{Goldstein} H.~Goldstein, {\it Classical Mechanics, 2nd ed.} (Addison
Wesley, Reading, MA, 1980), chap.~7, sec.~8.

\bibitem{string_theory} L.~Brink, S.~Deser, B.~Zumino, P.~Di Vecchia and
P.~Howe, Phys.~Lett.~B {\bf 64}, 435 (1976); S.~Deser and B.~Zumino,
Phys.~Lett.~B {\bf 65}, 369 (1976).

\bibitem{Tolman} R.C.~Tolman, {\it Relativity, Thermodynamics and Cosmology}
(Oxford University Press, London, 1934), chapter V, part II.

\bibitem{Weinberg} S.~Weinberg, {\it Gravitation and Cosmology} (Wiley,
New York, 1972), chapter 2, section 10.

\bibitem{MTW} C.W.~Misner, K.S.~Thorne and J.A.~Wheeler, {\it Gravitation}
(Freeman, San Francisco, 1973), chapter 22.

\bibitem{SW} R.L.~Seliger and G.B.~Whitham, Proc.~R.~Soc.~A {\bf 305}, 1 (1968).

\bibitem{Newcomb} W.A.~Newcomb, Nucl.~Fusion (1962 Suppl. part 2), 451 (1962).

\bibitem{tensor} This tensor also appears in the Lagrangian formulation of 
nonlinear photon-neutrino interactions; the effective Lagrangian given in 
\cite{photon_neutrino} has the form $G_{F} \alpha^{3/2}\, [a\,({\sf M}:{\sf F})
\,({\sf F}:{\sf F}) - b\,({\sf M}\cdot {\sf F}): ({\sf F}\cdot{\sf F})]$, where 
$a$ and $b$ are constants and $\alpha$ is the fine structure constant.

\bibitem{string_MHD} M.~Christensson and M.~Hindmarsh, Phys.~Rev.~D {\bf 60},
063001 (1999).
 
\bibitem{helicity_1} J.M.~Cornwall, Phys.~Rev.~D {\bf 56}, 6146 (1997).

\bibitem{helicity_2} M.~Giovannini, Phys.~Rev.~D {\bf 58}, 124027 (1998).

\bibitem{Kulsrud} R.M.~Kulsrud, R.~Cen, J.P.~Ostriker and D.~Ryu, Astrophys.~J.
{\bf 480}, 481 (1997).

\bibitem{neutron} It is true, however, that neutron-neutrino interactions can 
generate electromagnetic ($EM$) fields through the process $n \rightarrow \nu 
\rightarrow \sigma \rightarrow EM$; since this process is a second-order 
process (in powers of $G_{F}$), the electromagnetic fields thus produced are 
much smaller than the first-order fields considered here.

\bibitem{Moffatt_Ricca} H.K.~Moffatt and R.L.~Ricca, Proc.~R.~Soc.~Lond.~A
{\bf 439}, 411 (1992).

\bibitem{AMR} For more on Lie derivatives and applications of differential
geometry, see R.~Abraham, J.E.~Marsden and T.~Ratiu, {\it Manifolds, Tensor
Analysis, and Applications} (Addison-Wesley, Reading, Massachusetts, 1983),
chap.~6; or B.N.~Kuvshinov and T.J.~Schep, Phys.~Plasmas {\bf 4}, 537 (1997).

\end{thebibliography}
\end{document}